\documentclass[onecolumn,aps,preprint,showpacs,amsmath,amssymb]{revtex4-1}
\usepackage[pdftex]{graphicx}
\usepackage{latexsym}
\usepackage{dcolumn} 
\usepackage{color}
 
\usepackage{bm}

\begin{document} 


\newcommand{\eq}{\begin{equation}}
\newcommand{\eqe}{\end{equation}}

\title{Analytic solutions of the Madelung equation}

\author{Imre F. Barna$^{1,2}$ and Mih\'aly A. Pocsai$^{1,3}$ and L. M\'aty\'as$^{4}$}
\address{ $^1$ Wigner Research Centre of the Hungarian Academy of Sciences 
\\ Konkoly-Thege Mikl\'os \'ut 29 - 33, 1121 Budapest, Hungary \\
$^2$  ELI-HU Nonprofit Kft.,  Dugonics T\'er 13, H-6720 Szeged, Hungary \\
$^3$ University of P\'ecs, Institute of Physics, Ifj\'us\'ag \'utja 6 H-7624 P\'ecs, Hungary \\ 
$^4$Sapientia University, Department of Bioengineering, Libert\u{a}tii sq. 1, 530104 Miercurea Ciuc, Romania  }
\date{\today}
\begin{abstract}   We present analytic self-similar solutions 
for the one, two and three dimensional Madelung hydrodynamical equation for a free particle. There is a direct connection between the zeros of the  
Madelung fluid density and the magnitude of the quantum potential.  
\end{abstract}


\maketitle

\section{Introduction}
 
Finding classical physical basements of quantum mechanics is a great challenge since the advent of the theory. Madelung was one among the firsts who gave one explanation, this was the hydrodynamical foundation of the Schr\"odinger equation \cite{madelung1, madelung2}. 
His exponential transformation simply indicates that one can model quantum statistics hydrodynamically. The transformed equation has an attractive 
feature that the Planck's constant appears only once, as the coefficient of the quantum potential or pressure.  Thus, the fluid dynamicist can gather
experience of its effects by translating some of the elementary situations of the quantum theory into their corresponding fluid mechanical statements and vice versa. 

The quantum potential also appears in the de Broglie-Bohm pilot wave theory \cite{de-broglie,bohm1} (in other context) which is a non-mainstream attempt to 
interpret quantum mechanics as a deterministic non-local theory. In the case of  $\hbar \rightarrow 0 $ the Euler equation transforms to the Hamilton-Jacobi 
equation.

Later, many authors tried to generalize this Madelung's description for more compound quantum mechanical systems (like particles in external electromagnetic fields or with spin 1/2 ) \cite{janossy}.  	
Takabayashi \cite{taka} tried to interpret the Ansatz of Madelung as an ensemble of trajectories. 
Sch\"onberg \cite{schon} developed a new type of hydrodynamical model for the quantum mechanics for any values of spin. The quantum potential 
appears as a combination of a pressure term arising from the turbulence. 
  
In spite of the flaws of Madelung hydrodynamics (it cannot give a proper solution of the problem of atomic eigenstates and to the quantum description of emission or absorption processes), this approach turns to be fruitful in a number of applications like the stochastic quantum mechanics \cite{stoh}, quantum
cosmology \cite{cosm}, description of quantum-like systems \cite{systems}, the coherent properties of high-energy charged particle beams 
 \cite{beams1, beams2}.

Terlecki applied the fluid dynamical interpretation of the quantum mechanical 
probability density and current for the trajectory method and evaluated the solution of the time-dependent Schr\"odinger equation in atomic physics and calculated ionization and electron transfer cross sections for proton-hydrogen collision \cite{terl}.

As an interesting peculiarity Wallstrom showed with mathematical means that the initial-value problem of the Madelung equation is not well-defined and additional conditions are needed \cite{wallstrom}.  
 
Tsubota summarized the hydrodynamical descriptions of quantum condensed 
fluids such as superfluid helium and Bose-Einstein condensates as quantum hydrodynamics based on the original Ansatz of Madelung \cite{tsubota}. 

Nowadays, hydrodynamical description of quantum mechanical systems is 
a popular technical tool in numerical simulations. Review articles on quantum trajectories can be found in a booklet edited by Huges in 2011 \cite{qtraj}. 

From general concepts as the second law of thermodynamics a weakly non-local extension of ideal fluid dynamics can be derived which  
leads to the Schr\"odinger-Madelung equation as well \cite{van}. 

In the following study we investigate the Madelung equation with the self-similar 
Ansatz and present analytic solutions with discussion.  

This way of investigation is a powerful method to study the global properties of the solutions of various non-linear partial differential equations(PDEs) \cite{sedov}.  
Self-similar Ansatz describes the intermediate asymptotic of a problem: it is hold when the precise initial conditions are no longer important, but before the system has reached its final steady state. This is much more simpler than 
the full solutions and so easier to understand and study the different regions of the parameter space. A final reason for studying them is that they are solutions of a system of ordinary differential equations(ODEs) and hence do not suffer the extra inherent numerical problems of the full PDEs. In some cases self-similar solutions help to understand diffusion-like properties or the existence of compact supports of the solution. 

In the last years we successfully applied the multi-dimensional generalization of the self-similar Ansatz to numerous viscous fluid equations \cite{imre1,imre2} ending up with a book chapter of \cite{imre_book}.  
 
To our knowledge there are no direct analytic solutions available for the Madelung equation. Baumann and Nonnenmacher \cite{baum} exhaustively investigated the Madelung equation with Lie transformations and presented numerous ODEs, however non exact and explicit solutions are presented in a transparent way.  
Additional numerous studies exist where the non-linear Schr\"odinger equation is investigated with the Madelung Ansatz ending up with solitary 
wave solutions, \cite{roman} however that is not the field of our present interest.
 
\section{Theory and results} 
 
Following the original paper of Madelung \cite{madelung2}  the time-dependent
Schr\"odinger equation reads 
\eq
\triangle  \Psi - \frac{8\pi^2 m}{h^2} U \Psi - 
i\frac{4\pi m}{h} \frac{\partial \Psi}{\partial t}  = 0, 
\label{schrod}
\eqe
where $\Psi, U,m,h$ are  the  wave function, potential, mass and Planck's constant, respectively. 
Taking the following Ansatz $\Psi = \sqrt{\rho} e^{i S}$
where $\rho(x,t)$ and $S(x,t)$ are time and space dependent functions. 
 Substituting this trial function into Eq. (\ref{schrod}) going through the derivations the real and the imaginary part give us the following continuity and Euler equations with the form of 
\begin{eqnarray}
\rho_t + \nabla \cdot (\rho {\bf{v}}) &=& 0, \nonumber \\
{\bf{v}}_t + {\bf{v}} \cdot \nabla  {\bf{v}} &=& \frac{h^2}{8 \pi^2 m^2} 
\nabla\left(  \frac{\triangle \sqrt{\rho}}{\sqrt{\rho}} \right) - \frac{1}{m}\nabla U,  
\label{mad_euler}
 \end{eqnarray}
with ${\bf{v}} = \frac{\hbar}{m} {\bf{\nabla}} S$. The $\rho$ is the density of the investigated fluid and  ${\bf{v}}$ is the velocity field. Madelung also showed that this is a rotation-free flow.  
 The transformed equations has an attractive 
feature that the Planck's constant appears only once, at the coefficient of the quantum potential or pressure, which is the first term of the right hand side of the second equation. 
Note, that these are most general vectorial equation for the velocity field ${\bf{v}}$ which means that one, two or three dimensional motions 
can be investigated as well. 
In the following we will consider the two dimensional flow motion  ${\bf{v}} = (u,v) $ in Cartesian coordinates without any external field $U = 0$.
The functional form of the three and one dimensional solutions will be mentioned  
briefly as well. 

We are looking for the solution of Eqs. (\ref{mad_euler})
with self-similar Ansatz which is well-known from \cite{sedov} 
\eq 
\rho(x,y,t)=t^{-\alpha}f\left(\frac{x+y}{t^\beta}\right):=t^{-\alpha}f(\eta),
\hspace*{5mm}  u(x,y,t)=t^{-\delta}g(\eta),  \hspace*{5mm} 
v(x,y,t)=t^{-\epsilon}h(\eta), 
\label{self}
\eqe 
where $f,g$ and $ h$ are the shape functions of the density and the velocity field, respectively. 
The similarity exponents $\alpha, \beta, \delta, \epsilon $ are of primary physical importance since $\alpha, \delta, \epsilon $  represents the rate of decay of the magnitude of the shape function while $\beta$ represents the
spreading.  More about the general properties of the Ansatz can be found in 
our former papers \cite{imre1,imre2}.  Except some pathological cases 
all positive similarity exponents mean physically relevant dispersive solutions with decaying features at $x,y,t  \rightarrow \infty $.   
    Substituting the Ansatz (\ref{self}) into (\ref{mad_euler}) 
and going through some algebraic manipulation the following ODE system can be 
obtained for the shape functions
\begin{eqnarray}
-\frac{1}{2}f - \frac{1}{2}f'\eta + f'g + fg' + f'h + fh' = 0, \nonumber \\ 
-\frac{1}{2}g - \frac{1}{2}g'\eta + gg' + hg' = 
\frac{\hbar^2}{2m^2}\left(\frac{f'^3}{2f^3} -  \frac{f'f''}{f^2} +
  \frac{f'''}{2f}\right),  \nonumber \\ 
-\frac{1}{2}h - \frac{1}{2}h'\eta  + gh' + hh' =  
 \frac{\hbar^2}{2m^2}\left(\frac{f'^3}{2f^3} -  \frac{f'f''}{f^2} +
  \frac{f'''}{2f}\right).
\end{eqnarray}
Note that the particle mass appears in the denominator of the quantum potential term which is consistent with the experience of regular quantum mechanics that quantum features are relevant at small particle masses. 

The first continuity equation can be integrated giving us the mass as a conserved quantity and the parallel solution for the velocity fields $\eta = 2(g+h) + c_0$ 
where $c_0$ is the usual integration constant, which we set to zero. 
(A non-zero $c_0$ remains an additive constant in the final ODE (\ref{dens})  
as well.)   
It is interesting, and unusual (in our practice) that even the Euler equation can be 
integrated once giving us an other constant of motion. 
For classical fluids this is not the case.    
After some additional algebraic steps a decoupled ODE can be derived for the shape function of the density 
\eq
2f''f - (f')^2 + \frac{m^2 \eta^2f^2}{2\hbar^2} = 0. 
\label{dens}
\eqe
All the similarity exponents have the fixed value of $ +\frac{1}{2}$ which 
is usual for regular heat conduction, diffusion or for Navier-Stokes equations 
\cite{imre_book}. 
Note, that the two remaining free parameters are the mass of the particle $m$ and $\hbar$ which is the Planck's constant divided by $2\pi$. 
For a better transparency we fix $\hbar = 1$. 

At this point it is worth to mention, that the obtained ODE for the density shape function is very similar to Eq. (\ref{dens}) for different space dimensions, the only difference is a constant in the last term. For one, two or three dimensions the denominator has a factor of 1,2 or 3, respectively. 

Some additional space dependent potentials $U$ (like a dipole, or harmonic oscillator interaction) in the original Schr\"odinger equation would generate an extra fourth term in Eq. (\ref{dens}) like $\eta, f(\eta)$, or $\eta^2$. Unfortunately, no other analytic closed form solutions can be found for such ODEs.   
 
The solution of (\ref{dens}) can be expressed with the help of the Bessel functions of the first and second kind \cite{NIST} 
and has the following form of 
\eq
f(\eta) = \frac{2 \left(-J_{\frac{1}{4}} \left[\frac{\sqrt {2} m\eta^2}{8} \right]\cdot c_1 +  Y_{\frac{1}{4}} \left[\frac{\sqrt{2}m\eta^2}{8}\right] \cdot c_2  \right)^2}
{\eta^3 m^2 \left( J_{-\frac{3}{4}} \left[ \frac{\sqrt{2}m\eta^2}{8} \right] \cdot 
Y_{\frac{1}{4}} \left[ \frac{\sqrt{2}m\eta^2}{8} \right]   
- J_{\frac{1}{4}} \left[ \frac{\sqrt{2}m\eta^2}{8} \right] \cdot 
Y_{-\frac{3}{4}} \left[ \frac{\sqrt{2}m\eta^2}{8} \right]
\right)^2}
\label{bess}
\eqe
where $c_1$ and $c_2$ are the usual integration constants. 
The correctness of this solutions can be easily verified via back substitution 
into the original ODE.

To imagine the complexity of these solutions 
Figure 1 presents $f(\eta)$ for various $m$ values. 
It has a strong decay with a stronger and stronger oscillation at large arguments. 
The function is positive for all values of the argument, (which is physical for a 
fluid density), but such oscillatory profiles are completely unknown in regular fluid mechanics \cite{imre_book}. The most interesting feature is the infinite number of zero values which cannot be interpreted physically for a classical real fluid. 
 
\begin{center}
\vspace*{0.3cm}
  \resizebox{10cm}{!}
{   \rotatebox{0} 
    {
     \includegraphics{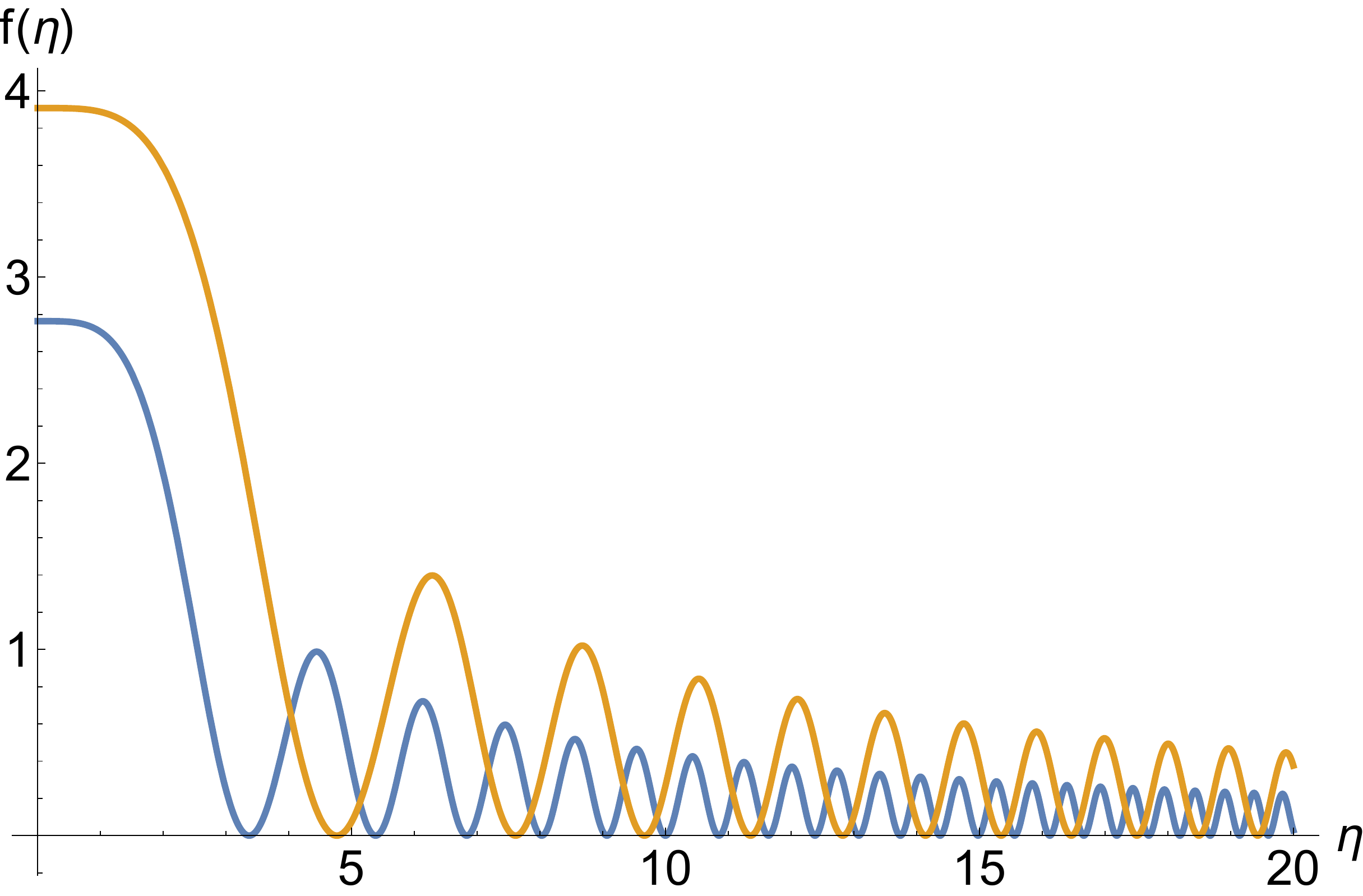}
      }}  \\
   {\bf{Fig. 1.} } The solution of Eq. (\ref{bess})  ($c_1=c_2=1$) the yellow curve is for m =1 and blue curve is for m= 0.5.
\label{F1}     
\end{center}
\begin{center}
\vspace*{0.3cm}
  \resizebox{10cm}{!}
{   \rotatebox{0} 
    {
     \includegraphics{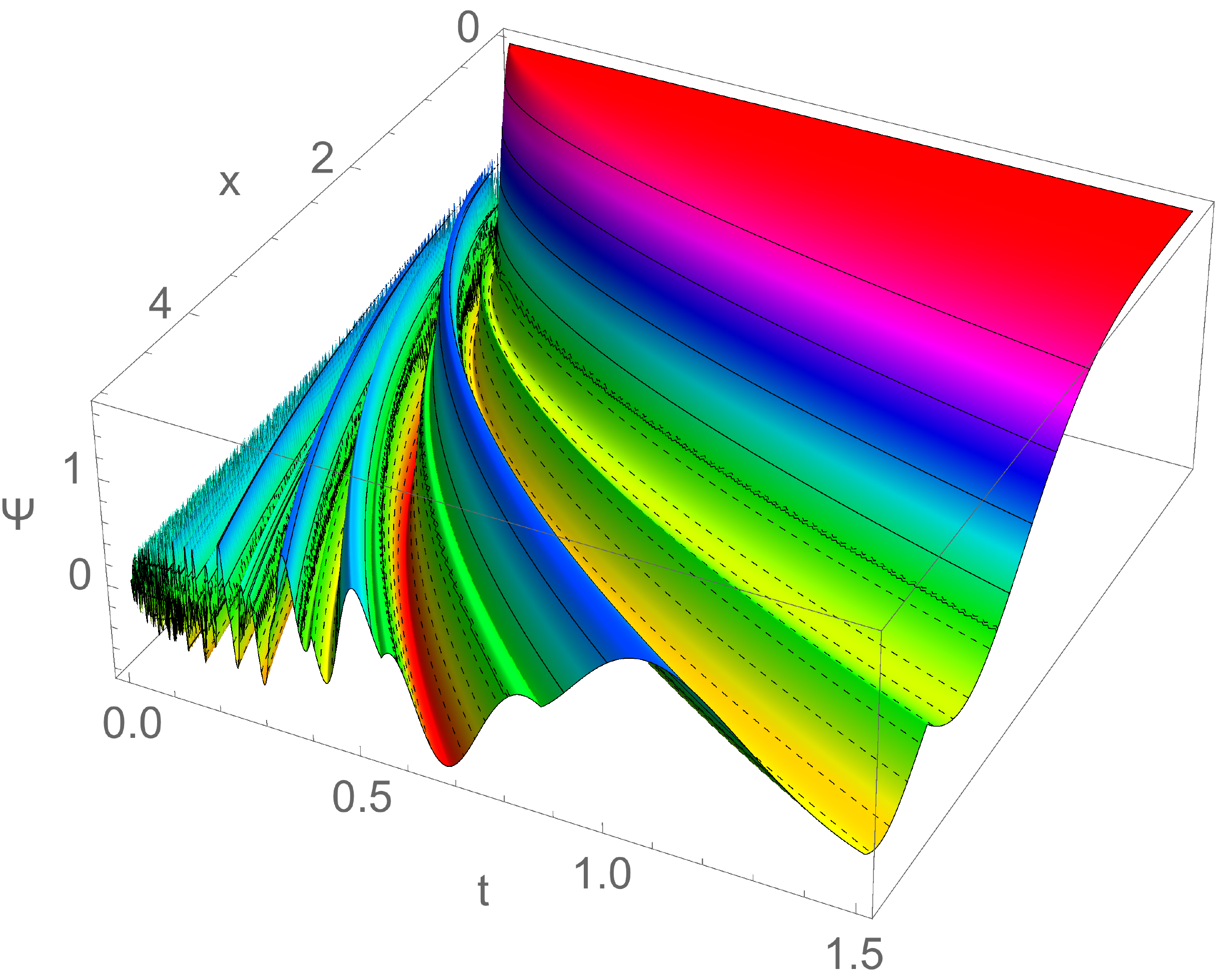}
      }}  \\
   {\bf{Fig. 2.} } The projection of real part of the wave function $\Psi(x,t)$ from Eq. (\ref{wavef}) for $m=1=c_1=c_2$. 
\label{F2}
\end{center}
The presented  form of the shape function cannot be simplified further, only $Y_{\nu}$s can be expresses with the help of $J_{\nu}$s \cite{NIST}.
Applying the recurrence formulas the orders of the Bessel functions can be shifted as well. 
With the parabolic cylinder functions the Bessel functions with $3/4$ and $1/4$ orders can be expressed, too. Unfortunately, all these formulas and manipulations are completely useless now. 

Both $J_{\nu}$ and $Y_{\nu}$ Bessel functions with linear argument form an orthonormal set, therefore integrable over the $L^2$ space.  
In our case, the integral $ \int_0^{\infty }f(\eta) d\eta$ is finite as well, unfortunately can be evaluated only by numerical means.  
This is good news because $f(\eta)$ is the density
function of the original Schrödinger equation.  In this sense $\sqrt{f}$ is the fluid mechanical analogue of the
real part of the wave function of the free quantum mechanical particle which can be described with a Gaussian wave packet.  
To obtain the complete original wave function, the imaginary part has to be evaluated as well. It is trivial from $\eta = \frac{x+y}{t^{1/2}}= 2(g+h)$
that 
\eq
 S = \frac{m}{\hbar}\int_{\bf{r}_0}^{\bf{r}_1}  {\bf{v}}{\bf{dr}} = 
\frac{m}{\hbar}\frac{(x+y)^2}{4t}. 
\eqe 

Now 
\begin{equation}
\begin{split}
 \Psi(x,y,t) =& \frac{ \sqrt{2}t^{\frac{1}{4}} \left(-J_{\frac{1}{4}} \left[\frac{\sqrt {2} m \{x+y\}^2}{8t} \right]\cdot c_1 +  Y_{\frac{1}{4}} \left[\frac{\sqrt{2}m\{x+y\}^2}{8t}\right] \cdot c_2  \right)}
{ ( x+y)^{3/2} m \left( J_{-\frac{3}{4}} \left[ \frac{\sqrt{2}m \{x+y \}^2}{8t} \right] \cdot 
Y_{\frac{1}{4}} \left[ \frac{\sqrt{2}m \{x+y \}^2}{8t} \right]   
- J_{\frac{1}{4}} \left[ \frac{\sqrt{2}m \{x+y \}^2}{8t} \right] \cdot 
Y_{-\frac{3}{4}} \left[ \frac{\sqrt{2}m \{x+y \}^2}{8t} \right]
\right)} \cdot \\ 
  & e^{\frac{mi}{\hbar} \frac{(x+y)^2}{4t}}\label{wavef}
\end{split}
\end{equation}
Figure 2 shows the projection of the real part  wave function to the ${x,t}$ sub-space.  
At small times the oscillations are clear to see, however at larger times the strong damping is evident. 

For arbitrary quantum systems, the wave function can be evaluated 
according to the Schr\"odinger equation, however we never know directly how large is the quantum contribution to the classical one.  
Now, it is possible for a free particle to get this contribution. (The Schr\"odinger equation gives the Gaussian wave function for a freely propagating particle.)  With the Madelung Ansatz we got the classical fluid dynamical analogue of the motion with the physical parameters $\rho(x,y,t), {\bf{v}}(x,y,t)$ which can be calculated analytically via the self-similar Ansatz thereafter original wave function $\Psi(x,y,t)$ of the quantum problem can be evaluated as well. 
The magnitude of the quantum potential $Q$ directly informs us where quantum effects are relevant. 
This can be evaluated from the classical density of the Madelung equation (\ref{mad_euler}) via 
\eq
Q = \frac{h^2}{8 \pi^2 m^2} 
\nabla\left(  \frac{\triangle \sqrt{\rho}}{\sqrt{\rho}} \right) = \frac{h^2}{8 \pi^2 m^2} \frac{\partial}{\partial \eta} 
\left( \frac{-\eta^2 m^2}{c_1 8 J_{\frac{1}{4}} [\frac{m \eta^2}{4\sqrt{2}} ]   -c_28 Y_{\frac{1}{4}} [ \frac{m \eta^2}{4\sqrt{2}} ]   }  \right). 
\label{quant_pot}
\eqe 
Figure 3 shows the shape function of quantum potential $Q(\eta)$ comparing to the shape function of the density $f(\eta)$. Note, that 
where the density has zeros the quantum potential is singular.  Such singular potentials might appear in quantum mechanics, however 
the corresponding wave function should compensate this effect. 
This is the main message of our study. Unfortunately, we cannot squarely state, that this kind of property is true in general for all the
Madelung quantum potentials (e.g.~for other Ans\"atze). Therefore additional investigations should be made to clarify this conjecture. 
\begin{center}
\vspace*{0.3cm}
  \resizebox{10cm}{!}
{   \rotatebox{0} 
    {
     \includegraphics{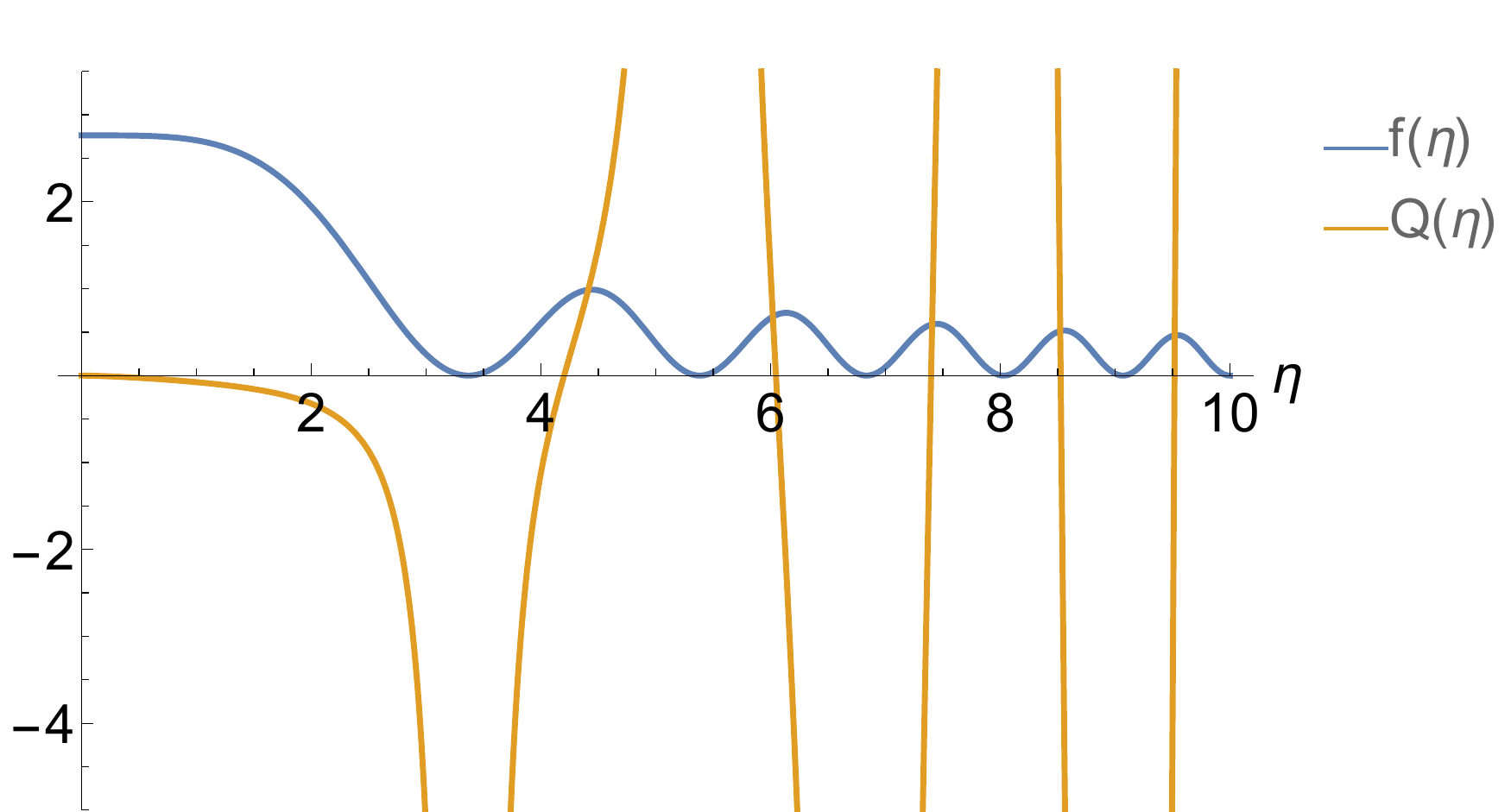}
      }}  \\
   {\bf{Fig. 3.} } The shape function of the density $f(\eta)$ is the blue curve  and the shape function of the quantum potential $Q(\eta)$ is the yellow 
solid line. All the corresponding parameters are $c_1 = c_2 = m =1$. 
\label{F3} 
\end{center}
\section{Summary and Outlook}
After reviewing the historical development and interpretation of the Madelung equation we introduced the self-similar Ansatz which is a not-so-well-known 
but powerful tool to investigate non-linear PDEs.  The free particle Madelung equation was investigated in two dimensions with this method, 
(the one and three dimensional solutions were mentioned as well.) We found  analytic solution for the fluid density, velocity field and the original wave function.
All can be expressed in terms of the Bessel functions.  
The classical fluid density has interesting properties, oscillates and has infinite number of zeros which is quite unusual and not yet been seen in such analysis.     




\end{document}